# Excitonic optical interface for GHz–THz collective excitations in a van der Waals magnet

Sophie Bork[a*], Richard Leven[a], Vincent Wirsdörfer[a], Alessandro Ferretti[b], Rafael R. Rojas-Lopez[a], Mattia Benini[a], David Maximilian Janas[a], Umut Parlak[c], Alberto Brambilla[b], Alexey V. Scherbakov[a], Swagata Acharya[d], Mirko Cinchetti[a*]

[a]*Department of Physics, TU Dortmund University, Dortmund, Germany*

[b]*Dipartimento di Fisica, Politecnico di Milano, 20133 Milan, Italy*

[c]*Fachbereich Physik, Universität Konstanz, Konstanz, 78457 Germany*

[d]*National Laboratory of the Rockies (NLR), 15313 Denver West Parkway, Golden, Colorado 80401, USA*

**Corresponding authors: sophie.bork@tu-dortmund.de, mirko.cinchetti@tu-dortmund.de*

## Abstract

Collective spin and lattice excitations in quantum materials span energy scales from GHz to THz, yet establishing a unified optical interface for these modes remains a central challenge. Here we show that excitonic resonances in the van der Waals antiferromagnet CrSBr provide a broadband optical interface for such excitations. Using femtosecond broadband transient reflectivity, we resolve coherent GHz magnon and THz phonon modes that modulate the dielectric response over a wide spectral range. Despite their distinct microscopic origin and frequency scales, both excitations give rise to the same emergent optical signature: a resonance at 1.46 eV that is absent in steady-state spectra and exhibits a characteristic π-phase inversion, identifying it as a discrete excitonic transition. We attribute this behaviour to boson-driven modulation of the dielectric response, which transiently transfers spectral weight from a nominally dark exciton into an observable channel without requiring a finite equilibrium oscillator strength. Supported by many-body calculations, we assign this feature to a higher-energy exciton with distinct momentum and orbital character and strongly suppressed optical matrix elements. These results establish excitonic resonances in van der Waals magnets as a platform for interfacing collective excitations across GHz, THz and optical frequency scales.

## Introduction

The optical response of quantum materials is governed by excitons—bound electron–hole pairs whose energy, symmetry and coherence encode the underlying electronic structure[1–3]. Beyond their role in equilibrium spectroscopy, excitonic resonances provide a potential interface between optical degrees of freedom and lower-energy collective excitations. In many materials, these excitations span widely separated frequency scales, from GHz to THz, including spin, lattice and hybrid spin–lattice modes, yet establishing a unified pathway to couple them to optical observables remains a central challenge[4–8]. While optically bright excitons dominate steady-state spectroscopy, a large fraction of the excitonic manifold remains hidden[9–11]. These so-called dark excitons, suppressed by symmetry or orbital selection rules, play a central role in energy relaxation[12], transport[13] and coherence dynamics[14] but remain largely inaccessible to conventional probes. Developing routes to access and couple these hidden states is therefore key to linking collective excitations across disparate energy scales within a common optical framework.

Materials that can host such an interface must combine strong excitonic effects with well-defined and coherent low-energy collective excitations across multiple frequency scales. Van der Waals antiferromagnets provide an ideal platform in this regard. In these systems, strong Coulomb interactions stabilize robust excitons, while long-range magnetic order and lattice dynamics give rise to coherent spin and phonon excitations spanning GHz to THz frequencies[15–17]. The semiconductor CrSBr[18–20], which exhibits long-range A-type antiferromagnetic order below $T_N$ = 132 K[21–24], combines a highly anisotropic electronic structure[25–27] with well-defined near-infrared excitonic resonances, including two prominent bright excitons at 1.36 eV and 1.77 eV[5,28–31], and supports both GHz acoustic magnons[4,32,33] and THz optical phonons[34–37]. Recent studies have identified an additional excitonic state near 1.5 eV with strongly suppressed oscillator strength[38], pointing to the presence of a dark exciton whose optical signatures remain elusive.

Here we show that excitonic resonances in the van der Waals antiferromagnet CrSBr provide a broadband optical interface for collective excitations. Using broadband femtosecond transient reflectivity, we resolve coherent GHz magnon and THz phonon dynamics that modulate the dielectric response over a wide spectral range. The resulting modulation is not restricted to a specific excitonic state but extends across the excitonic manifold, encompassing both optically bright and nominally dark excitons. Despite their distinct microscopic origin and widely separated frequency scales, both excitations give rise to the same emergent optical signature: a resonance at 1.46 eV that is absent in steady-state spectra but exhibits the characteristic phase response of a discrete excitonic transition. We demonstrate that this behaviour arises from boson-driven modulation of the electronic response, which transiently renders an optically suppressed exciton observable. These results establish excitonic resonances in van der Waals magnets as a platform for coupling collective excitations across GHz, THz and optical frequency scales.

## Results and discussion

We begin by examining the time-resolved optical response of CrSBr in the spectral range encompassing its low-energy excitonic resonances. Broadband transient reflectivity reveals a pronounced coherent oscillatory signal extending across the entire probe window (1.25–1.65 eV), indicative of a coherent collective response. To disentangle the different dynamical contributions, measurements were performed over distinct temporal windows and sampling intervals chosen to independently resolve GHz and THz dynamics (see Methods). This approach enables us to separately capture slow GHz oscillations associated with coherent magnons and faster THz oscillations arising from optical phonons.

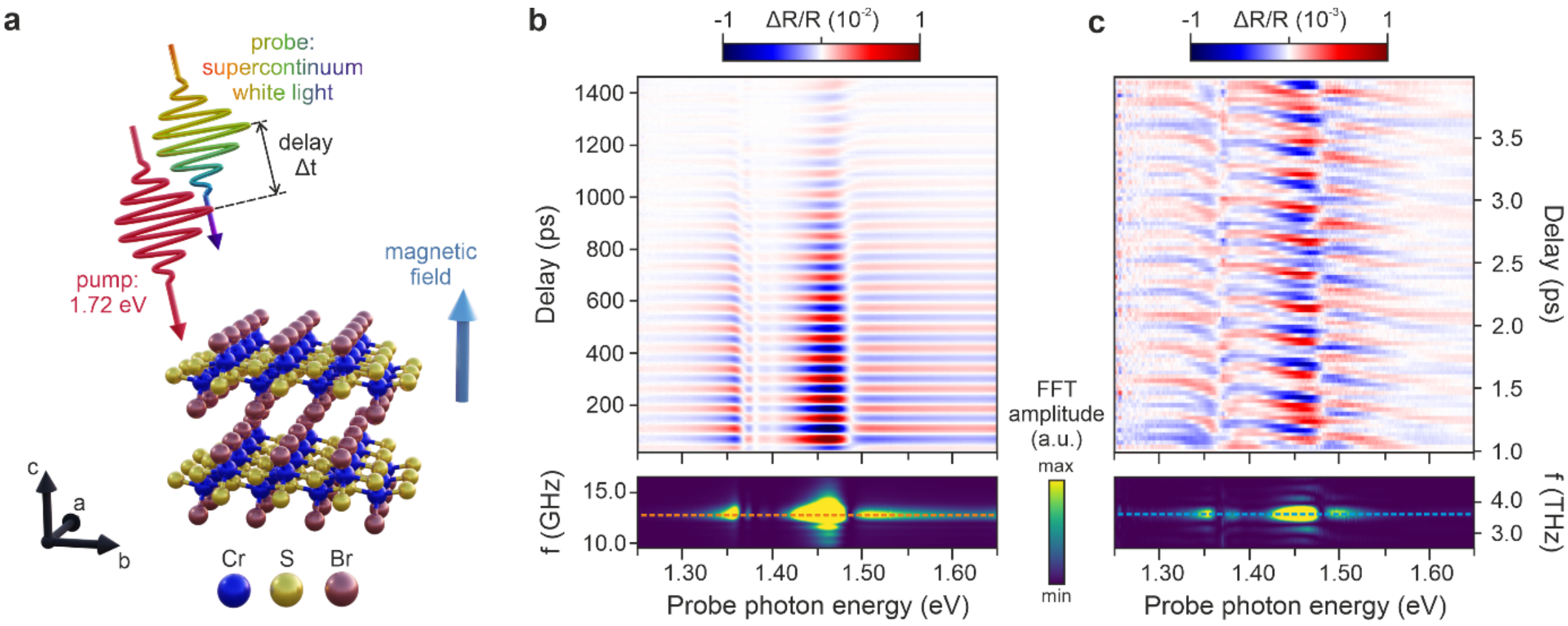

*Figure 1: Experimental scheme and transient reflectivity response of CrSBr across GHz and THz timescales. a) Schematic of the measurement geometry, showing stacked CrSBr layers, the applied magnetic field (B ∥ c), and near-normal-incidence pump and probe beams. b) Top: Transient reflectivity (ΔR/R) at B = 0.7 T for pump-probe delays from 25 to 1475 ps after subtraction of the incoherent background, revealing clear GHz oscillations across 1.25–1.65 eV with enhanced amplitude near 1.36 eV and 1.46 eV. Bottom: Two-dimensional map of the fast Fourier transform (FFT) of each spectral channel in b), showing a dominant mode at 12.74 GHz (dashed orange line) with intensity maxima near 1.36 and 1.46 eV. c) Top: Same as b), but for pump-probe delays from 1 to 4 ps, revealing pronounced THz oscillations around 1.36 and 1.46 eV. Bottom: Two-dimensional FFT map of c), displaying a dominant mode of 3.63 THz (dashed blue line).*

**Fig. 1a** illustrates the experimental configuration for the transient reflectivity measurements on bulk CrSBr at 80 K, with an out-of-plane magnetic field applied parallel to the crystallographic c-axis. Pump and probe beams were incident at near-normal angles, with the pump photon energy set to 1.72 eV, close to resonance with the higher-energy bright exciton at 1.77 eV, and the probe spanning a broadband supercontinuum from 1.25 to 1.65 eV. Pump and probe polarizations were aligned along the crystallographic a and b axes, respectively, and transient reflectivity dynamics were recorded for magnetic fields up to 2.4 T at a pump fluence of 2.88 mJ/cm$^2$.

The resulting transient response separates naturally into two distinct frequency regimes. **Fig. 1b** shows the low-frequency response measured over pump–probe delays from 25 ps to 1475 ps after subtraction of the incoherent background (see SI). Clear oscillations are observed across the full probe-energy range, and Fourier analysis reveals a dominant mode at 12.74 GHz with negligible probe-energy dispersion. **Fig. 1c** displays the high-frequency response measured over delays from 1 ps to 4 ps after background subtraction (see SI). In this regime, Fourier analysis identifies a dominant oscillation at 3.63 THz.

Despite the nearly three orders of magnitude difference in frequency and their distinct microscopic origin, both modes exhibit remarkably similar spectral fingerprints. In particular, the FFT amplitude maps of both the GHz and THz responses display enhanced spectral weight near 1.36 eV and 1.46 eV. The appearance of the same spectral features in both frequency regimes immediately suggests that coherent spin and lattice excitations couple to a common excitonic structure in CrSBr, revealing a common excitonic optical response.

We first consider the low-frequency response shown in **Fig. 1b**. To identify the origin of the 12.74 GHz mode, we performed magnetic-field-dependent measurements over the range B = 0–2.4 T. As shown in the Supplementary Information, the oscillation frequency evolves systematically with magnetic field and the mode intensity strongly decreases outside the field range between 0.2 and 1.1 T. In particular, the oscillatory response is absent at B = 0 T and vanishes above B = 1.2 T. These characteristics identify the mode as a coherent magnon excitation, consistent with previous reports of GHz spin dynamics in CrSBr[4,33].

We next turn to the high-frequency response shown in **Fig. 1c**. In addition to the dominant mode at 3.63 THz, a weaker feature at 7.61 THz is observed (see SI), consistent with optical phonon modes reported previously[34,37]. Because the 7.61 THz oscillation lies close to the experimental resolution limit, we focus on the 3.63 THz mode in the following. In contrast to the GHz response, the 3.63 THz oscillation exhibits negligible magnetic-field dependence over the entire investigated range from 0 to 2.4 T. Moreover, the mode remains clearly visible both at zero field and above 1.2 T, where the GHz oscillation disappears. These observations identify the high-frequency response as a coherent optical phonon mode.

Having established the microscopic origin of the two oscillatory responses, we compare their spectral dependence. **Figs. 1b** and **1c** reveal that both the GHz magnon and THz phonon modes produce enhanced FFT amplitudes at the same probe photon energies near 1.36 and 1.46 eV. Notably, these resonances are observed only for probe polarization aligned along the crystallographic b-axis, whereas no corresponding features are detected for probe polarization along a (see SI). This pronounced polarization selectivity reflects the strongly anisotropic electronic structure of CrSBr[39,40] and is consistent with the predominantly b-polarized optical response of its low-energy excitonic manifold reported previously[28,41–44]. The emergence of identical spectral fingerprints from two collective excitations separated by nearly three orders of magnitude in frequency demonstrates that both modes probe the same excitonic structure in CrSBr.

To analyse the origin of the spectral features observed in the GHz and THz responses, we extract line profiles of the FFT amplitude at the respective mode frequencies (along the dashed lines in **Figs. 1b** and **1c**, respectively). As shown in **Fig. 2a**, both the magnon and phonon responses exhibit two pronounced resonances centered near 1.36 eV and 1.46 eV. The lower-energy feature coincides with the well-established bright 1s exciton of CrSBr[5,28–31], whereas the higher-energy resonance appears in a spectral region where no corresponding feature is observed in steady-state optical spectroscopy.

To verify the excitonic character of these resonances, we examine the phase behaviour of the coherent oscillations as a function of probe photon energy. **Fig. 2b** shows transient reflectivity traces of the GHz oscillations measured across the high-energy resonance near 1.46 eV. As the probe energy is tuned across the trailing edge of the feature, the oscillatory response undergoes a clear π-phase inversion. An analogous phase reversal is observed for the THz response around the same spectral position after bandpass filtering to suppress the incoherent background and the higher-frequency phonon contribution (**Fig. 2c**).

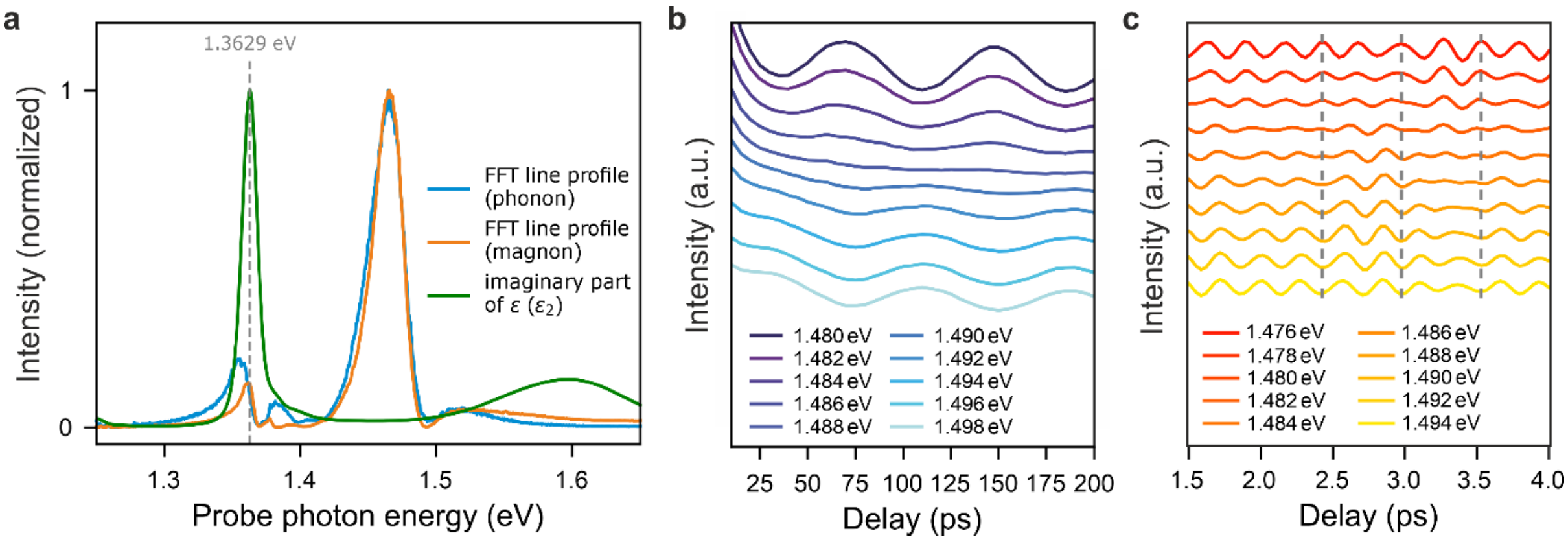

***Figure 2: Excitonic fingerprints in the optical response of CrSBr. a)*** *Comparison between transient and static optical responses at B = 0.2 T (all spectra normalized). The orange and blue curves show the FFT amplitudes extracted from the GHz magnon and THz phonon responses, respectively, while the green curve shows the imaginary part of the dielectric function,* $\varepsilon_2$*, obtained from static reflectivity measurements via Kramers–Kronig analysis. All spectra exhibit a pronounced resonance near 1.36 eV corresponding to the bright 1s exciton. In contrast, the resonance near 1.46 eV appears only in the transient response, consistent with a nominally dark excitonic state with strongly suppressed equilibrium oscillator strength.* ***b)*** *Phase-resolved transient reflectivity traces of the GHz-oscillations measured across the high-energy resonance near 1.46 eV at B = 0.7 T (traces vertically offset). A clear π-phase inversion is observed at the trailing edge of the resonance, indicating the presence of a discrete electronic transition.* ***c)*** *Same as in b), but for THz oscillatory response around the 1.46 eV resonance after bandpass filtering to suppress the incoherent background and the higher-frequency phonon contribution. The dashed grey lines highlight the π-phase reversal.*

This phase inversion provides a stringent fingerprint of a discrete excitonic transition. In transient reflectivity, coherent bosonic oscillations periodically modulate the dielectric function, producing an oscillatory component in ΔR/R proportional to the spectral derivative of the optical response. Across a discrete resonance, this derivative changes sign, leading to a reversal of the oscillation phase. The observation of a π-phase shift at both 1.36 eV and 1.46 eV therefore demonstrates that the corresponding spectral features originate from well-defined excitonic resonances.

While the resonance at 1.36 eV is readily identified with the bright 1s exciton of CrSBr[45], the origin of the 1.46 eV feature is less straightforward. Previous studies reported photoluminescence resonances above 1.45 eV[34], as well as a weak photoreflectance feature attributed to higher-lying electronic transitions[46]. More recently, resonant inelastic x-ray scattering identified a dark excitonic excitation near 1.46 eV[38]. The observation of a robust π-phase inversion at this energy now establishes that the transient optical response couples to a discrete excitonic state despite the absence of a corresponding resonance in equilibrium optical spectra.

To further clarify the nature of the 1.46 eV resonance, we compare the transient response with steady-state optical measurements. **Fig. 2a** shows the imaginary part of the dielectric function $\varepsilon_2$ obtained from static reflectivity measurements via Kramers–Kronig analysis (green curve; see SI), together with the FFT amplitudes extracted from the GHz magnon and THz phonon responses (orange and blue curves, respectively). A pronounced resonance is observed in all three spectra near 1.36 eV, consistent with the bright 1s exciton of CrSBr. In contrast, no corresponding feature is present in the static dielectric response near 1.46 eV, despite the clear resonance observed in both the GHz and THz transient spectra. The absence of a steady-state optical signature demonstrates that this excitonic state possesses a strongly suppressed equilibrium oscillator strength and becomes detectable only in the transient response. The fact that both magnons and phonons reveal the same hidden resonance further indicates that its visibility does not rely on a specific bosonic excitation, but instead emerges from a more general mechanism of coherent boson-driven optical modulation. Together with the observed π-phase inversion and recent reports of a dark excitonic excitation near 1.5 eV[38], these findings strongly support the assignment of the 1.46 eV feature to a nominally dark exciton in CrSBr. In this picture, coherent bosonic dynamics transiently render the optically suppressed exciton observable in the time-resolved response.

To gain microscopic insight into the origin of the hidden excitonic resonance and its visibility in the transient optical response, we performed many-body calculations of the excitonic spectrum of CrSBr using self-consistent Feynman diagram-based approach QSG$\hat{W}$[47–49] (see Methods). The calculations clearly identify a higher-energy excitonic state near 1.50 eV with strongly suppressed oscillator strength, consistent with the experimentally observed resonance at 1.46 eV. In contrast to the bright exciton at 1.36 eV, which is dominated by transitions near the Brillouin-zone center Γ, the higher-energy state is primarily constructed from electronic states near the Brillouin-zone boundary close to X (**Fig. 3a**). This demonstrates that a q = 0 exciton in CrSBr does not need to originate from Γ, but can emerge from other regions of the Brillouin zone provided that the electron and hole share the same crystal momentum. The two excitons also exhibit markedly different orbital character. While the bright 1.36 eV exciton is dominated by hole states of predominantly $d_{yz}$ character[44,50,51] the higher-energy dark exciton near 1.50 eV is primarily associated with $d_{xy}$-derived states (**Fig. 3b**). This distinction is physically significant in CrSBr because the strongly anisotropic crystal structure stabilizes the $d_{yz}$ low-energy sector through the elongated b-axis bonding geometry, thereby favoring optically active low-energy excitons. By contrast, the dark exciton involves a less favorable orbital channel together with a more localized real-space character, providing a natural microscopic explanation for its strongly suppressed optical matrix elements.

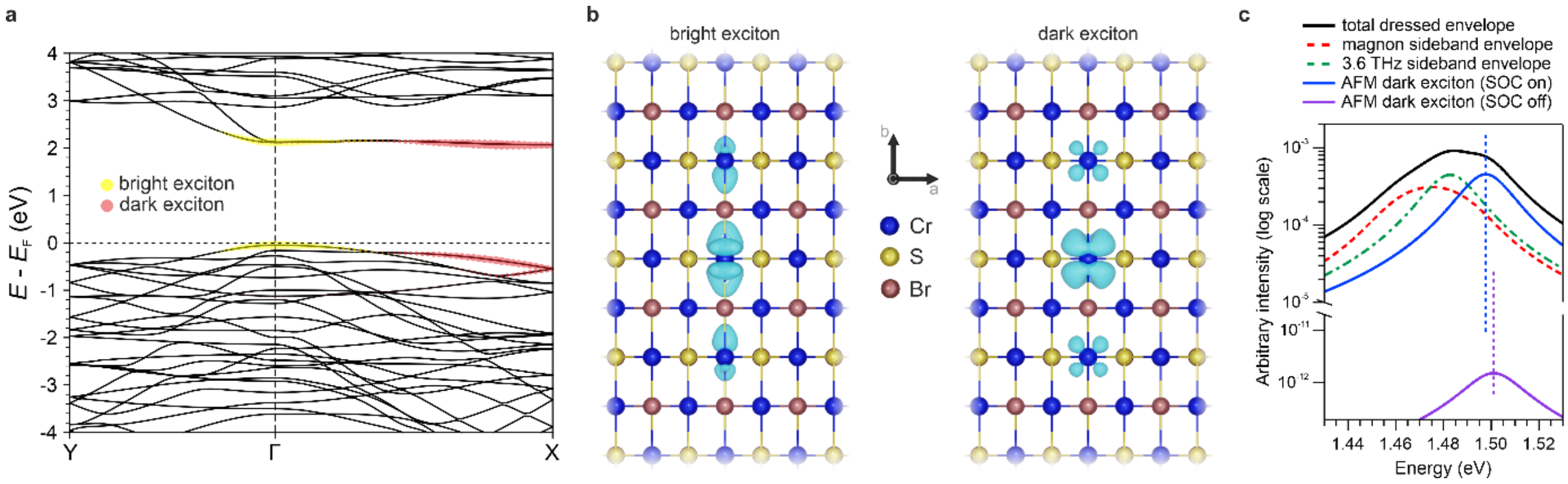

***Figure 3: Microscopic origin and boson-dressed optical response of the dark exciton in CrSBr. a)*** *Projection of the excitonic wavefunctions onto the electronic band structure obtained from the QSGŴ calculations. The bright exciton near 1.36 eV is dominated by transitions close to the Brillouin-zone center Γ, whereas the higher-energy dark exciton near 1.50 eV primarily originates from states near the Brillouin-zone boundary close to X.* ***b)*** *Real-space and orbital decomposition of the bright and dark excitons. The dark exciton exhibits a more localized character, enhanced onsite orbital contributions, and a distinct hole sector compared with the bright exciton.* ***c)*** *Calculated oscillator strength of the dark exciton with (blue) and without (purple) spin-orbit coupling (SOC). While SOC leaves the exciton energy nearly unchanged, it weakly relaxes the optical selection rules and increases the oscillator strength by many orders of magnitude, thereby rendering the otherwise optically suppressed exciton weakly allowed. Black, red, and green curves: boson-dressed spectral envelopes constructed using experimentally motivated magnon and phonon energy scales. Coherent bosonic excitations redistribute spectral weight from the parent dark exciton toward lower energies, generating transient optical signatures below the original exciton energy. The experimentally observed redistribution energies are consistent with the characteristic magnon and phonon energy scales.*

The calculations further reveal substantial differences in the orbital character of the two excitons. Compared to the bright exciton, the higher-energy state exhibits enhanced dipole-forbidden onsite dd character and reduced intersite dipole-allowed dd contributions, resulting in strongly suppressed optical matrix elements (see SI). Importantly, the dark exciton is already present in the antiferromagnetic QSGŴ calculation without spin-orbit coupling, where it appears at 1.50 eV with an essentially vanishing oscillator strength along the b-axis ($f_{010} \sim 10^{-14}$). Including spin-orbit coupling leaves the exciton energy nearly unchanged but increases the oscillator strength by several orders of magnitude ($f_{010} \sim 10^{-5}$) (**Fig. 3c**). This demonstrates that spin-orbit coupling does not create a different excitonic state, but instead relaxes the optical selection rules of an already existing parent exciton.

Starting from this weakly allowed parent exciton, we then construct boson-dressed spectral envelopes using experimentally motivated magnon and phonon energy scales (**Fig. 3c**). In this framework, coherent bosonic excitations redistribute spectral weight from the parent dark exciton toward lower energies, generating transient optical signatures below the original exciton energy. Consistent with this picture, both the coherent phonon and magnon channels can modulate the optical visibility of the dark-exciton manifold and redistribute spectral weight within the experimentally relevant energy window. While the present model is not intended to quantitatively reproduce the transient spectra, it supports the interpretation that coherent bosonic dynamics provide a natural route for dressing and activating the hidden excitonic response.

Within this picture, coherent magnons and phonons act as periodic perturbations of the dielectric response that dynamically probe the same hidden excitonic manifold. The observation that both GHz and THz excitations produce identical optical fingerprints therefore does not arise from direct hybridization between excitons and bosons, but from a common boson-driven modulation of the excitonic response. These results establish excitonic resonances in van der Waals magnets as a broadband optical interface linking spin, lattice and optical degrees of freedom across widely separated frequency scales.

**Conclusion**

In conclusion, we have demonstrated that excitonic resonances in the van der Waals antiferromagnet CrSBr provide a common optical response to collective excitations spanning GHz and THz frequency scales. Coherent magnons and phonons generate the same emergent optical fingerprint, revealing a hidden excitonic state near 1.46 eV that remains invisible in equilibrium optical spectroscopy. Supported by many-body calculations, we identify this feature as a nominally dark exciton with strongly suppressed oscillator strength and distinct momentum and orbital character. Our results establish excitonic resonances in van der Waals magnets as a platform for interfacing spin, lattice and optical degrees of freedom across multiple energy scales.

## Methods

**<u>Transient reflectivity setup</u>**

Transient reflectivity measurements were carried out using an Yb-based femtosecond laser system (PHAROS, Light Conversion) operating at a central wavelength of 1040 nm and an average output power of 20 W[16,52]. The output beam was divided into pump and probe arms with powers of 13 W and 7 W, respectively. Pump and probe photon energies were tuned independently using optical parametric amplifiers (OPAs): the pump was set to 1.72 eV, while the probe was tuned to 1.09 eV. A broadband supercontinuum probe was generated by attenuating the probe beam with a variable neutral-density filter and focusing it onto a 3-mm sapphire plate. After transmission through a short-pass filter (FESH1000, Thorlabs), the usable spectral window extended from 1.25 to 1.65 eV. The pump–probe delay was introduced by a motorized delay line (Newport) positioned in the pump path. To resolve GHz and THz oscillations, delay steps of 5 ps and 0.033 ps were used, respectively. Pump and probe beams were each routed through a half-wave plate and a polarizer, enabling independent control of polarization and power. Both beams were then focused onto the sample surface using spherical mirrors. The reflected probe beam was collected, collimated, and spectrally dispersed by a diffraction grating, and subsequently detected on a calibrated multichannel diode array (Stresing), providing spectrally resolved transient-reflectivity signals with high sensitivity.

The sample was mounted inside a cryostat capable of reaching temperatures down to 4 K. An external magnetic field of up to 2.4 T was applied out of plane using a superconducting solenoid integrated around the cryostat. Both pump and probe beams were directed onto the sample under (near-)normal incidence.

**<u>Static reflectivity setup</u>**

For the static reflectivity measurements, we used a broadband supercontinuum generated by a static white-light source (Thorlabs SLS201L/M). The beam polarization was controlled by a linear polarizer. All measurements were performed under normal incidence in the spectral range from 1.25 to 2.75 eV. A microscope objective (Mitutoyo 50× NIR) was used to focus the beam onto the sample, resulting in a spot size of approximately 10 µm. The reflected light was first imaged onto a camera to determine and adjust the measurement position on the sample. By flipping a mirror out of the optical path, the reflected light was then directed to a compact spectrometer (Thorlabs CCS200/M). For data analysis, the measured spectra were normalized to the measured spectrum of the white-light source.

## Sample

A custom-grown CrSBr crystal (HQ Graphene) with lateral dimensions of approximately 8 × 2 mm (crystallographic *a*- and *b*-axis) was used for the measurements. Immediately prior to the experiments, the crystal surface was freshly cleaved using scotch tape, resulting in a final bulk thickness of about 2 µm (crystallographic *c*-axis). Sample quality was assessed through Raman spectroscopy measurements (see Supplementary Information).

## Many body calculations of the excitonic spectrum

The excitonic spectrum of bulk CrSBr was computed within self-consistent many-body perturbation theory using the quasiparticle self-consistent GW framework augmented by ladder diagrams in the screened interaction, i.e. the QSG$\hat{W}$ implementation described in detail in Ref. [47–49]. In this approach, quasi-particle self-consistent GW – QSGW[53,54] removes the starting-point dependence of conventional $G_{DFT}W_{DFT}$, while the inclusion of ladder diagrams in W accounts for electron-hole interactions in the polarizability and substantially improves both the quasiparticle gap and the optical response. QSG$\hat{W}$ is a self-consistent extension of QSGW. Within QSG$\hat{W}$, the electronic eigenfunctions are computed in the presence of screened Coulomb correlations corrected by an excitonic vertex. Self-consistency is imposed for both the self-energy, $\Sigma$, and the charge density. It has been shown to be particularly important when magnetic degrees of freedom are involved, as there is an additional coupling between spin and $\Sigma$. For bulk CrSBr in the antiferromagnetic phase, a 12-atom unit cell was used with lattice parameters a = 3.504 Å and b = 4.738 Å. The one-particle Hamiltonian and dynamical self-energy $\Sigma(k,\omega)$ were constructed on a 10×7×2 k-mesh. QSGW and QSG$\hat{W}$ cycles were iterated until the RMS change in the static quasiparticlized self-energy $\Sigma_0$ fell below $10^{-5}$ Ry. The two-particle Hamiltonian entering the Bethe-Salpeter equation contained 26 valence bands and 9 conduction bands, and excitonic eigenvalues and eigenfunctions were converged on a 10×7×2 k-mesh too by diagonalizing a two-particle Hamiltonian of rank 32760.

To analyse the hidden exciton relevant for the transient experiment, excitonic eigenvalues, oscillator strengths, band-basis projections, and real-space/orbital decompositions were extracted from the BSE solutions. In the without SOC calculation, the dark exciton appears at 1.50 eV with essentially vanishing oscillator strength along the b axis, $f_{010} = 5.7\times 10^{-14}$. Recomputing the BSE bubble in the presence of spin-orbit coupling leaves the exciton energy nearly unchanged (1.497 eV) but increases its oscillator strength to $f_{010} = 1.7 \times 10^{-5}$, demonstrating that SOC relaxes the optical selection rule of an already existing parent exciton rather than creating a qualitatively different state.

To illustrate how coherent bosons can redistribute spectral weight below this parent exciton, we constructed separate magnon- and phonon-dressed spectral envelopes starting from the AFM+SOC dark exciton line. The parent exciton was represented as $A_0(\omega) = f_d\, L(\omega; E_d, \Gamma_d)$, where E_d = 1.497 eV is the dark-exciton energy, $f_d = 1.7 \times 10^{-5}$ is its oscillator strength along the b axis, and $L(\omega; E_d, \Gamma_d)$ is a normalized Lorentzian centered at $E_d$ with width $\Gamma_d$, introduced only as a visualization broadening of the discrete excitonic state. The magnon-dressed envelope was computed as $A_m(\omega) = \eta_m \int d\Omega\, D_m(\Omega)\, A_0(\omega + \Omega)$, where $D_m(\Omega)$ is the antiferromagnetic magnon density of states and $\eta_m$ is an overall coupling prefactor that controls the weight of the magnon sideband. The phonon-dressed envelope was computed analogously as $A_{ph}(\omega) = \eta_{ph} \int d\Omega\, D_{ph}(\Omega)\, A_0(\omega + \Omega)$, where $D_{ph}(\Omega)$ is the phonon spectral function, taken here as a narrow Gaussian centered at the coherent optical phonon frequency 3.63 THz (approximately 14.9 meV) observed experimentally, and $\eta_{ph}$ is the corresponding phonon-coupling prefactor. This construction should be viewed as a physically motivated sideband-envelope model that illustrates how bosonic dressing redistributes spectral weight below the parent dark exciton, rather than as a quantitative fit to the transient spectra.

**AI usage**

After completion of the initial manuscript, the text was further refined regarding readability and clarity with the assistance of large language models. All suggested improvements were carefully reviewed by the authors and incorporated when deemed appropriate. No additional use of artificial intelligence was made during data acquisition, analysis, interpretation, or other aspects of this study.

## Acknowledgments

This work was supported by the Deutsche Forschungsgemeinschaft (DFG) through the Project dd2D (Project number 555818086.) and by the European Union's Horizon 2020 Research and Innovation Programme under Project SINFONIA, grant 964396. We gratefully acknowledge Cristian Manzoni (IFN-CNR) for his valuable support and advice during the setup and optimization of the transient reflectivity apparatus. This work was authored in part by the National Laboratory of the Rockies for the U.S. Department of Energy (DOE) under Contract No. DE-AC36-08GO28308. For S.A., funding was provided by the Computational Chemical Sciences program within the Office of Basic Energy Sciences, U.S. Department of Energy. S.A. acknowledges the use computational resources sponsored by the Department of Energy's Office of Energy Efficiency and Renewable Energy and located at the National Laboratory of the Rockies. The views expressed in the article do not necessarily represent the views of the DOE or the U.S. Government. The U.S. Government retains and the publisher, by accepting the article for publication, acknowledges that the U.S. Government retains a nonexclusive, paid-up, irrevocable, worldwide license to publish or reproduce the published form of this work, or allow others to do so, for U.S. Government purposes.